\documentclass[runningheads]{llncs}
\usepackage{graphicx}
\usepackage{times}
\usepackage{helvet}
\usepackage{courier}
\usepackage{booktabs}
\usepackage{amssymb,amsmath,amsfonts}   
\usepackage{bm}
\usepackage{algorithm}
\usepackage{algorithmic}
\usepackage{xcolor}
\graphicspath{{./figures/}}
\DeclareGraphicsExtensions{.pdf,.jpeg,.png}

\newcommand{\richard}[1]{\textcolor{black}{#1}}
\newcommand{\eepeng}[1]{\textcolor{black}{#1}}

\newcommand{\commentout}[1]{}
\begin{document}
\title{JobComposer: Career Path Optimization via Multicriteria Utility Learning \thanks{This research is supported by the National Research Foundation, Prime Minister's Office, Singapore under its International Research Centres in Singapore Funding Initiative.}}
%
%
\author{Richard J. Oentaryo \and
Xavier Jayaraj Siddarth Ashok \and
Ee-Peng Lim \and 
Philips Kokoh Prasetyo
}
\authorrunning{R. J. Oentaryo et al.}
%
\institute{Mclaren Applied Technologies, 7 Temasek Boulevard, Singapore 038987 \\
\email{richard.oentaryo@mclaren.com} \and
Living Analytics Research Centre, Singapore Management University, Singapore 178902
\email{\{xaviera,eplim,pprasetyo\}@smu.edu.sg}\\
}
\maketitle              
\begin{abstract}
With online professional network platforms (OPNs, e.g., LinkedIn, Xing, etc.) becoming popular on the web, people are now turning to these platforms to create and share their professional profiles, to connect with others who share similar professional aspirations and to explore new career opportunities.  These platforms however do not offer a long-term roadmap to guide career progression and improve workforce employability.  The career trajectories of OPN users can serve as a reference but they are not always optimal.  A career plan can also be devised through consultation with career coaches, whose knowledge may however be limited to a few industries. To address the above limitations, we present a novel data-driven approach dubbed \texttt{JobComposer} to automate career path planning and optimization. Its key premise is that the observed career trajectories in OPNs may not necessarily be optimal, and can be improved by learning to maximize the sum of payoffs attainable by following a career path. At its heart, \texttt{JobComposer} features a decomposition-based multicriteria utility learning procedure to achieve the best tradeoff among different payoff criteria in career path planning. Extensive studies using a city state-based OPN dataset demonstrate that \texttt{JobComposer} returns career paths better than other baseline methods and the actual career paths.

\end{abstract}
\section{Introduction}
\label{sec:introduction}

\textbf{Motivation.} At some point in our life, we may raise questions such as ``What is my career goal?'' or ``How can I achieve my career goal?'' 
Developing a career plan is essential for the career success of individuals, giving them a sense of direction to follow and a way to track how their career is progressing. {Career planning would also help workforce improve their employability and develop the right skills, especially in the face of uncertainties due to rapid shifts in global economy and technology.

The success of career planning depends on how much insights one has on the set of possible career trajectories.
To gain such insights, one may conduct his or her own research or seek consultation and counseling from some career coach(es).
Both approaches are however challenging as they have to cope with dynamically-changing job and workforce landscape. 
To offer individuals career planning insights at the societal scale, there is a critical need to automate career planning as much as possible, thus motivating this research. 


Fortunately, with the fast-rising adoption of online professional network (OPN) sites (such as LinkedIn, Xing and Viadeo), people start creating and sharing their professional profiles with others, using their profiles to connect with others as well as to explore new career opportunities.  It is also possible to leverage on these OPNs to gain deeper and more comprehensive insights on career trajectories. 

\textbf{Objectives.} In this paper, we therefore develop a new data-driven approach dubbed \texttt{JobComposer} for automated career path planning and optimization. The centerpiece of this approach is a multicriteria utility learning procedure that seeks to identify an optimal strategy for composing a career path emanating from a given origin job, by taking into account the trade-off among different payoff criteria simultaneously. Our goal is to assist job seekers and career coaches in making more informed career decisions and guidance, respectively, based on a data-centric view of job and workforce landscape.  

To the best of our knowledge, the problem of optimizing (improving) existing career paths is new, and our \texttt{JobComposer} is the first attempt to address this problem via utility learning. Previous works on OPN-based career path recommendation \cite{Lou2010,Liu:AAAI2016} favor the more popular career paths that people take, implying an inherent assumption that common career paths are ideal. However, this is not necessarily true in reality, and so the utility values of these popular career paths may be suboptimal (see Section~\ref{sec:experiments}).

\texttt{JobComposer} works based on the premise that observed career paths may be suboptimal and can be further improved. The suboptimality can be due to the ``incompleteness'' of observed career paths as their users may discontinue further career for various reasons.  For example, a user could decide to retire at an early age, or fail to acquire the required skills to advance to the next higher level job.  Another possible reason of suboptimality is the subjectivity in career decisions made by users.  For instance, an head of engineering who is no longer interested in senior management, may relinquish her position to become an engineer. Finally, the optimality of a career path involves multiple criteria, e.g., salary, job level, etc..
To realize this, \texttt{JobComposer} assembles career paths out of all possible job transitions found in observed career trajectories, and employs a new multicriteria utility learning  procedure that jointly takes into account multiple payoff criteria when optimizing a career path. This allows us to capture the tradeoff among different---possibly competing---goals in career path planning. As individuals select their preferred criteria, \texttt{JobComposer} would return the corresponding optimized career paths. However, these career paths are not personalized as the preferred criteria differ from user features (e.g. skills, education level, etc.). 


\textbf{Contributions}. We summarize our main contributions as follows: (a) We present a new problem formulation of career path optimization as a multicriteria utility learning task using career trajectories observed in OPN data, which distinguishes our work from previous works \cite{Wang:WWW2013,Liu:AAAI2016} that do not consider a set of payoff criteria in finding good career paths; (b) We develop an efficient decomposition-based iterative procedure that divides the multicriteria utility learning problem into multiple scalar optimization subproblems and optimizes them simultaneously, which has been shown to be efficient than non-decomposition-based methods; and (c) We conduct extensive experiments to evaluate the efficacy and validity of our approach in both quantitative and qualitative aspects.


\textbf{Paper outline.} In Section \ref{sec:related_work}, we first survey related works on job transition analyses and career recommendation. Section~\ref{sec:dataset} describes our dataset.  We then elaborate the proposed \texttt{JobComposer} approach in Section \ref{sec:method}. Section \ref{sec:experiments} presents the details of our empirical study using a city state-based OPN dataset. Section \ref{sec:conclusion} concludes this paper. 
\section{Related Work}
\label{sec:related_work}

Research on job and workforce movements has been around for decades \cite{NBERw2649,LabourMobility,moscarini2007occupational,fuller2008job,Joseph:2012}, which often involves analyzing various aspects of workforce such as age and wage growth.  
These studies, however, traditionally relied on surveys, census, and other data such as tax lists and population registers, which require time-consuming and costly (manual) efforts to collect. Moreover, the findings tend to be focused on selected workforce segments, and cannot be easily replicated in a larger population.

With the rise of OPN, there is now a growing interest to conduct research on online user data in order to understand job transition behavior and career growth. 
Wang \emph{et al.} \cite{Wang:WWW2013} proposed a hierarchical Bayesian model to predict the probability of a user making a job transition at a certain time.
Cheng \emph{et al.} \cite{Cheng:KDD2013} modeled job transition activities to rank influential companies.  
State \emph{et al.} \cite{State:Socinfo2014} analyzed the migration trends of professional workers into the US. 
Xu \emph{et al.} \cite{Xu:ICDM2015} combined work experiences from OPNs and check-in records from location-based social networks to predict job change occasions. 
More recently, Chaudhury \emph{et al.} \cite{Chaudhury:WWW2016} analyzed the growth patterns of the ego-network of new employees in companies.
Xu \emph{et al.} \cite{Xu:KDD2016} generated and analyzed job transition networks to identify talent circles. Kapur \emph{et al.} \cite{Kapur:KDD2016} employed the PageRank \cite{Page1999} algorithm on a graph of job transitions among companies to estimate the desirability ranking of the companies, and use this ranking to compute the ranking of universities. 

Beyond job transition analysis, a few studies have been conducted that make use of the career paths (trajectories) of OPN users. Xu \emph{et al.} \cite{Xu:KDD2014} developed a sequence alignment method for quantifying the professional similarity between two career paths. Lou \emph{et al.} \cite{Lou2010}  devised a career path planning method that utilizes the shortest path algorithm on a job hop graph, where the edge weight is the negative logarithm of the job transition probability. Most recently, Liu \emph{et al.} \cite{Liu:AAAI2016} developed a multi-source learning framework that combines information from multiple social networks (i.e., sources) and models the career path of an individual. 

While our work is most closely related to \cite{Lou2010,Liu:AAAI2016}, it is noteworthy that the proposed \texttt{JobComposer} approach is based on a fundamentally different premise. In particular, the methods in \cite{Lou2010,Liu:AAAI2016} were designed to obtain a good fit to common/popular career trajectories observed in the OPN data. As such, they inherently assume that the common trajectories are ideal, which may not always hold in reality. 
In contrast, \texttt{JobComposer} works based on the premise that the observed career trajectories may be suboptimal and have room for improvements. \texttt{JobComposer} also goes further by incorporating multicriteria utility-based learning and path generation features. As such, it is able to achieve a good balance among multiple goals involved in a career path planning process.

\section{Dataset}
\label{sec:dataset}

\textbf{Dataset construction.} We study career trajectories data collected from one of the largest OPNs in the city state of Singapore.  We first crawled the directory of all public user accounts maintained by the target OPN.  
The dataset was constructed in November 2016.  These public user profiles cover professional attributes of users including their education and career trajectory information.   \eepeng{Each career trajectory consists of the sequence of jobs of a user and the respective companies.  Companies come with different sizes and are assigned to the size categories: [2-10], [11-50], [51-200], [201-1000], [1001-5000], [5001-10000], and [10001+]. Each company also belongs to an industry. }    

\textbf{Data cleaning.}  A job here refers to a job title in a certain sized company of an industry, as opposed to specific job title in some company.  This allows the resultant career paths to be more appropriate for guiding users as opposed to specific job in specific company.
\richard{To facilitate a proper empirical study, we perform two data cleaning steps. Firstly, we only include those jobs $s$, whose job titles $j_s$ have a minimum support of $100$ users. This removes jobs with unusual and misspelled titles. 
Secondly, we remove users who did not specify their education qualifications. We also remove all work experiences before one graduates from university or colleges.} Table~\ref{tab:basic_stats} summarizes the count statistics of the final dataset that we obtain after cleaning, and use throughout our study.

\textbf{Out-transition degree of jobs.} To determine the likelihood of out-transition of jobs in our dataset, we plot the out-transition degree of jobs in Figure~\ref{fig:hop_freq} against the career paths observed.  From the figure, we can observe that there are jobs with an out-degree of 0.  These represent the last jobs in the career paths. These are usually the CEO and other very senior jobs.  There are also many more jobs with out-degree greater than 1. 

\begin{table}[!t]
\scriptsize
\centering
\caption{Statistics of our OPN dataset}
\label{tab:basic_stats}
\begin{tabular}{|l|c|c|}
\hline
\scriptsize
\textbf{Count metric} & \textbf{Value} \\
\hline
No. of user profiles 	       	& 455,477\\
No. of career trajectories     	& 57,784\\
No. of (unique) jobs    		& 255,691\\
No. of (unique) job transitions & 265,533\\
\hline
\end{tabular}
\end{table}

\begin{figure}[!t]
\centering
\includegraphics[width=0.65\textwidth]{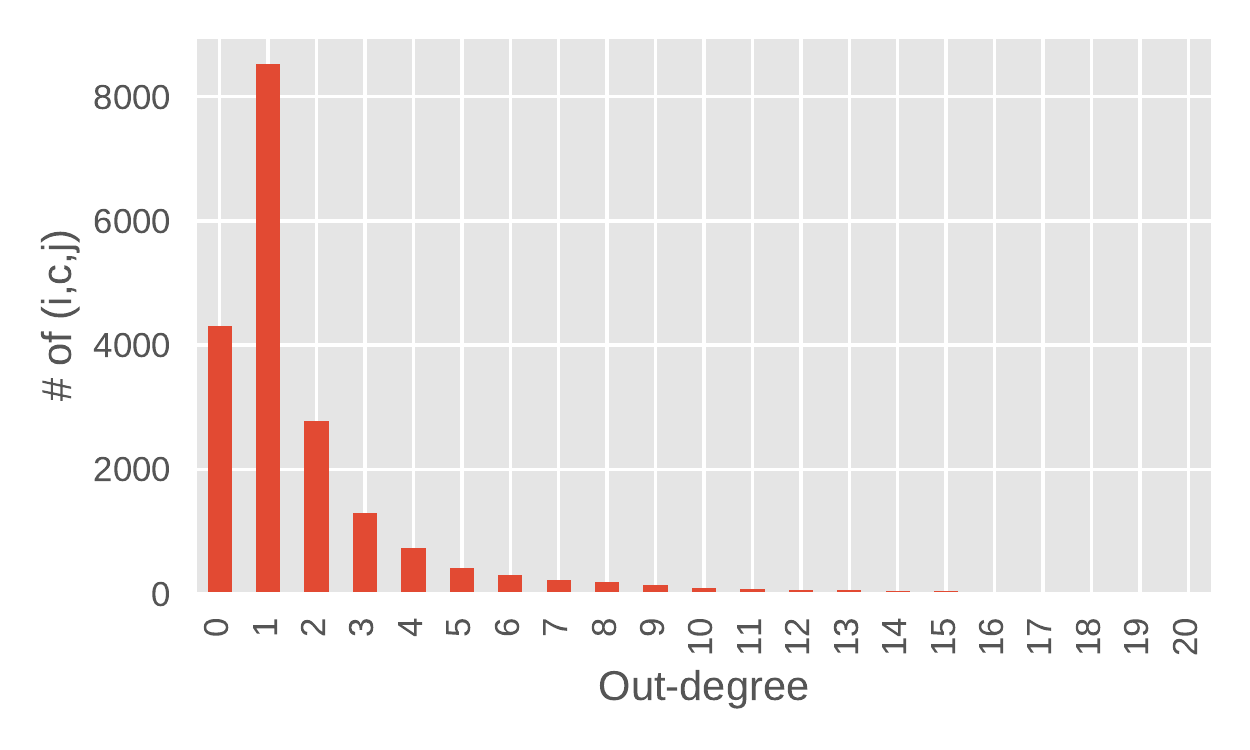}
\caption{Distribution of out-degree of jobs}
\label{fig:hop_freq}
\end{figure}

\section{Proposed Approach}
\label{sec:method}

\richard{In this section, we first introduce our problem formulation that forms the basis of our \texttt{JobComposer} approach. Next, we describe the core ideas of \texttt{JobComposer}, namely: decomposition-based multicriteria utility learning and multicriteria utility selection. Finally, we present several payoff criteria used in the context of our OPN data.}

\subsection{Problem Formulation}
\label{sec:problem}

\richard{First of all, we define a \emph{career path} as a sequence of jobs taken by a person in a chronological order. Here, two adjacent/consecutive jobs in the sequence are assumed to have non-overlapping time period. Each \emph{job} $s$ is represented by a tuple $(i_s, c_s, j_s)$, which denotes a job title $j_s$ for a company size category $c_s$ in industry $i_s$. Note that each company size category $c_s$ represents the company type in terms of employee size, e.g., `1-50 employees' typically denotes a small startup, whereas `10,000+ employees' usually refers to a large, multi-national corporation (MNC)}.

Based on this definition, we formulate the following problem: ``\emph{Given a person's current (origin) job, how can we find an optimal career path that maximizes the total payoff possibly obtained by following that path}?'' \richard{Under this formulation, we then treat career path as a stochastic process that undergoes transitions from one job (state) to another job (state). For computational simplicity, we particularly focus on a type of stochastic processes called \emph{Markov process}, which exhibits a Markov property such that the next value of the process depends on the current value, but is conditionally independent of the previous values of the stochastic process \cite{Elson2012}. In the context of career path planning, this means that a person's future job is stochastically independent of his/her jobs in the past, given his/her current job.}

\richard{It is worth noting that our formulation can be viewed as a special instance of \emph{Markov Decision Process} (MDP) \cite{Puterman1994}. One key difference is that the MDP formulation additionally includes the concept of action, which allows an agent to make a choice/decision that would affect the evolution of the agent's states over time. However, in our career planning context, only one action exists for the transition of one state (job) to another, which is hopping from one job $j$ to another job $j'$. In this case, the next state is a deterministic (non-stochastic) function of the current state and action, and thus MDP reduces to a Markov process. Extended formulation of the current career planning problem to MDP is beyond the scope of this paper and would be left for future work.}


Given the Markov process formulation, how do we then define the optimality of a career path beginning from some origin/source job $s$?} Firstly, we consider a person who has to decide his/her path of destination jobs $\{ d_1,\ldots,d_t,\ldots,d_T\}$, in order to maximize the discounted sum of future payoffs $r(d_{t-1}, d_{t})$ that he/she can possibly get starting from $s$.
The optimal value for this maximization process is given by the \emph{value function}:
\begin{align}
\label{eqn:value_func}
V(s) &= \max_{ \{ d_1,\ldots,d_t,\ldots,d_T \} } \left[ r(s, d_1) + \gamma r(d_1, d_2) + \ldots + \gamma^{T-1} r(d_{T-1}, d_{T}) \right]
\end{align}
where $s = d_0$ and $\gamma \in (0, 1]$ is the discount factor. The discount factor controls how myopic (or far-sighted) the selection of a career path is. In an economic context where the payoff is the amount of dollars made, a discount factor $\gamma < 1$ can be interpreted in terms of the idea that a dollar today is worth more than a dollar tomorrow.


Subsequently, the formula (\ref{eqn:value_func}) can be rearranged to:
\begin{align}
V(s) &= \max_{d_1} \left[ r(s, d_1) + \max_{ \{ d_{2}, \ldots, d_{T} \} } \left[ \gamma r(d_1, d_2) + \ldots + \gamma^{T-1} r(d_{T-1}, d_{T}) \right] \right] \nonumber\\ 
&= \max_{d_1} \left[ r(s, d_1) + \gamma \max_{ \{ d_{2}, \ldots, d_{T} \} } \left[ r(d_1, d_2) + \ldots + \gamma^{T-2} r(d_{T-1}, d_{T}) \right] \right] \nonumber\\ 
&= \max_{d} \left[ r(s,d) + \gamma V(d) \right]
\end{align}
where we equate $d = d_1$ for notational simplicity. This recursive definition is also known as the \emph{Bellman equation}, which underpins dynamic programming (DP) theory \cite{Bellman1957}. Next, we define a \emph{utility function} $U(s,d)$ for transition from job $s$ to $d$ such that:
\begin{align}
V(s) = \max_d U(s,d)
\end{align}
This in turn leads to the following recursive formulation:
\begin{align}
\label{eqn:utility_func}
U(s,d) &= r(s,d) + \gamma V(d) \nonumber\\
       &= r(s,d) + \gamma \max_{d'} U(d,d')
\end{align}

\eepeng{In the recursive formulation,  $d'$ denotes the next destination job that maximizes the overall utility.}  To solve equation (\ref{eqn:utility_func}), a simple way is to turn the equality symbol ``$=$'' into an assignment operator ``$\leftarrow$'' and keep iterating the assignments. This leads to the following \emph{value iteration} procedure: For all job transitions $(s, d)$, where $s$ and $d$ are the source and destination jobs respectively, repeat until terminating criterion:
\begin{align}
\label{eqn:value_iter}
U(s,d) \leftarrow r(s,d) + \gamma \max_{d'} U(d,d')
\end{align}
The above asynchronous DP update procedure is guaranteed to converge to the optimal utility $U^*(s,d)$, independent from the initial value of $U(s,d)$. We can prove the convergence by showing that, for large iteration $k$, $\lim_{k \rightarrow \infty} U(s,d)^{(k)} = U^*(s,d)$. To this end, we first quantify the distance between $U(s,d)^{(k)}$ and the optimal $U^*(s,d)$ using $L_{\infty}$-norm, i.e., $|| U^{(k)} - U^*||_{\infty} = \max_{(s,d)} |U(s,d)^{(k)} - U(s,d)^*|$. We then let $\mathcal{T}$ be the backup operator such that $\mathcal{T}(U(s,d)) = r(s,d) + \gamma \max_{d'} U(d,d')$. With this, we can show that:
\begin{align}
\label{eqn:convergence}
|| U^{(k+1)} - U^*||_{\infty} &\leq || \mathcal{T}(U^{(k)}) - \mathcal{T}(U^*) ||_{\infty} 
\leq \gamma || U^{(k)} - U^* ||_{\infty} \leq \ldots \nonumber\\
&\leq \gamma^{k+1} ||U^{(0)} - U^* || \rightarrow 0
\end{align}




\subsection{Decomposition-Based Multicriteria Utility Learning}
\label{sec:multi_utility}

The value iteration described in (\ref{eqn:value_iter}) operates based on a single type of payoff criterion. In practical career planning, however, one may be interested in multiple payoff criteria, such as job level gain, duration of stay in a company, income gain, etc. To account for multiple career planning goals, we generalize the single-criterion approach described in (\ref{eqn:value_iter}) and propose a new multicriteria utility learning procedure.

\begin{algorithm}[!t]
\begin{algorithmic}
\REQUIRE
\STATE Set of all job transition observations $\mathcal{D} = \{ (s,d) \}$, \STATE Discount factor $\gamma \in (0,1]$,
\STATE Maximum iteration $T_{max}$,
\STATE Total number of weight combinations $N$
\ENSURE
\STATE Utility values $\{ U(s,d,\vec{\lambda}^j) | \forall (s,d) \in \mathcal{D}, \forall \vec{\lambda}^j \in \bm{\lambda} \}$
\\\hrulefill
\STATE \textbf{ /* Initialization */}
\STATE Generate all $N$ possible weight combinations $\bm{\lambda} = [ \vec{\lambda}^1, \ldots, \vec{\lambda}^j, \ldots, \vec{\lambda}^N]$
\STATE Set $U(s,d,\vec{\lambda}^j) \leftarrow 0$ for all weight combinations $\vec{\lambda}^j \in \bm{\lambda}$ and job transitions $(s,d) \in \mathcal{D}$
\STATE \textbf{ /* Utility update */ }
\REPEAT
    \FOR {each weight combination $\vec{\lambda}^j \in \bm{\lambda}$}
    	\FOR {each job transition $(s,d) \in \mathcal{D}$}
    		\STATE $r(s,d,\vec{\lambda}^j) \leftarrow \sum_{i=1}^M \lambda_i^j f_i(s,d)$
        	\STATE $U(s,d,\vec{\lambda}^j) \leftarrow r(s,d,\vec{\lambda}^j) + \gamma \max_{d'} U(d,d',\vec{\lambda}^j)$
		\ENDFOR
    \ENDFOR
\UNTIL maximum iteration $T_{max}$
\end{algorithmic}
\caption{Decomposition-Based Multicriteria Utility Learning (MUL/D)}
\label{algo:multicriteria_learning}
\end{algorithm}

To this end, we first decompose the multicriteria utility problem into a number of scalar optimization subproblems by considering a weighted combination of the different criteria. Let $\vec{\lambda} = [\lambda_1, \ldots, \lambda_i, \ldots, \lambda_M]^T$ denote a weight vector, such that $\lambda_i \geq 0, \forall i \in \{1,\ldots,m\}$ and $\sum_{i=1}^M \lambda_i = 1$, where $M$ is the total number of criteria. Accordingly, we extend the payoff and utility functions by adding a third dimension  that corresponds to a particular weight  vector $\vec{\lambda}$. That is, the value iteration formula now becomes:
\begin{align}
\label{eqn:multi_value_iter}
U(s,d,\vec{\lambda}) \leftarrow r(s,d,\vec{\lambda}) + \gamma \max_{d'} U(d,d',\vec{\lambda})
\end{align}
where the payoff function $r(s,d,\vec{\lambda})$ is defined by:
\begin{align}
\label{eqn:weighted_sum}
r(s,d,\vec{\lambda}) = \sum_{i=1}^M \lambda_i f_i(s,d)
\end{align}
where the payoff function $r(s,d,\vec{\lambda})$ is defined by:
 $f_i(s,d)$ is the $i^{th}$ payoff criterion for transition $(s,d)$. We   further elaborate the payoff criteria we have used in Section \ref{sec:criteria}.

The expression given in (\ref{eqn:weighted_sum}) corresponds to the so-called \emph{weighted summation} scalarization method \cite{Miettinen1999}. Given a fixed weight vector $\lambda$, $r(s,d,\vec{\lambda})$ is a linear combination of $f_i(s,d)$. The value iteration procedure for $U(s,d,\vec{\lambda})$ will thus retain the same convergence property as equation (\ref{eqn:convergence}). An empirical demonstration of the convergence trait of our multicriteria utility learning procedure is presented in Section \ref{sec:experiments}.

Algorithm \ref{algo:multicriteria_learning} summarizes our decomposition-based multicriteria utility learning procedure, or MUL/D in short. The original multicriteria utility career path optimization is broken down into multiple weighted criteria single utility optimization problems, each with a unique weight combination.
Here, the number of criteria is $M$. The total number of possible weight combinations $N$ is controlled by a positive (integer) parameter $H$. More precisely, $\vec{\lambda}^1, \ldots, \vec{\lambda}^j, \ldots, \vec{\lambda}^N$ are all the weight vectors in which each individual weight $\lambda_i^j$ ($1 \leq i \leq M-1$) takes a value from $\{ \frac{0}{H}, \frac{1}{H}, \ldots, \frac{H}{H} \}$.
The final weight component, $\lambda_M^j$, is determined by:
\begin{align}
\lambda_M^j = 1 - \sum_{k=1}^{M-1} \lambda_k^j
\end{align}

With this definition, the total number of possible weight vectors $N$ for $M$ different criteria is given by:
\begin{align}
N = (H+1)^{M-1} 
\end{align}
As an example, suppose we have three criteria (i.e., $M=3$) and $H = 10$.  In this case, the degree of freedom is $M-1=2$, owing to the constraint $\sum_{i=1}^M \lambda_i = 1$. That is, the weight for the last (third) criterion can be computed as $\lambda_3 = 1 - \lambda_1 - \lambda_2$.  The total number of weight vectors or weight combinations would therefore be $(H+1)^{M-1} = 11^2 = 121$. By adjusting $H$ alone, we can control both learning overhead and granularity of the optimal solutions.

\subsection{Multicriteria Utility Selection}

Once we learn through Algorithm \ref{algo:multicriteria_learning} the utility values $U(s,d,\vec{\lambda_j})$ for all possible weight vectors $\lambda_j \in \bm{\lambda}$, the next question is: \emph{How do we pick a weight vector $\lambda^*$ that gives the best tradeoff/balance among different criteria}? To answer this, we introduce the idea that the best tradeoff/balance is one that allows best overall improvement of actual observed paths taken by individuals.  With this, we present a new metric known as the \emph{product of improvement means} (PIM) defined as:
\begin{align}
\label{eqn:PIM}
\text{PIM} &= \min_{i} \left( \text{sgn}(\mu_i) \right) \times \prod_{i=1}^M |\mu_i|
\end{align}
where $\text{sgn}()$ is the sign function (i.e., $sgn(\mu)=+1$ when $\mu > 0$ and -1 otherwise), and $\mu_i$ is the mean of improvement of the optimized payoff criterion $i$ over to the actual payoff criterion $i$:
\begin{align}
\mu_i = \frac{1}{K} \sum_{k=1}^K &\left( \sum_{(s,d) \in \mathbf{P}^\text{optimized}(k, s_0)} f_i(s,d) \right. 
- \left. \sum_{(s,d) \in \mathbf{P}^\text{actual}(k, s_0)} f_i(s,d) \right)
\end{align}
Here $\mathbf{P}^\text{optimized}(k, s_0)$ and $\mathbf{P}^\text{actual}(k, s_0)$ are the set of all job transitions in the optimized career path produced by \texttt{JobComposer} and those in the actual path taken by an individual $k$, respectively. In this case, both paths start from the same origin job $s_0$. The distinction between the two paths will be further illustrated in Section~\ref{sec:experiments}.

The term $\min_{i} ( \text{sgn}(\mu_i) )$ in (\ref{eqn:PIM}) implies that, if any one of $\mu_i$ is negative, then the overall PIM value will be negative as well. This is to ensure that each improvement mean $\mu_i$ has a positive value, which is intuitive and desirable. Without this constraint, it is possible that two of the criteria give negative improvement means, but the overall PIM is positive due to double-negation of the (negative) signs of the corresponding $\mu_i$ during the product operation.

For a given weight vector $\lambda^j$, we have a PIM value $\text{PIM}(\lambda^j)$. Accordingly, we identify the best tradeoff among different payoff criteria by picking a weight vector $\lambda^*$ such that the following condition is met:
\begin{align}
\label{eqn:PIM_sign}
 \lambda^* = \arg \max_{\lambda^j \in \bm{\lambda}} \text{PIM}(\lambda^j)
\end{align}

\subsection{Payoff Criteria}
\label{sec:criteria}

In this work, we devise a few examples of payoff criterion $f_i(s,d)$, which tell us how rewarding a job transition is.  These are payoff criteria that are derived from the career trajectories observed in our OPN dataset (which we will elaborate later). Our multicriteria utility learning approach is however not limited to these criteria.  Depending on the available datasets, one could use criteria such as salary, medical benefits, etc.. Let $p$ and $s$ (or $d$) denote a person and a job, respectively. Our three criterion examples are defined as follows:
\begin{itemize}
\item \textbf{Duration cost} ($D_{s,d}$): This is our simplest payoff criterion; for a job pair $(s,d)$, it refers to the \emph{average time lapse} between the start date and end date of the source job $s$, over all people who move from job $s$ to job $d$.  \eepeng{In other words, a job transition which takes less time duration to the destination job is viewed as more favorable than another job transition which takes more time duration.} Formally, we define the duration cost as:
\begin{align}
\label{eqn:duration}
D_{s,d} = \frac{\sum_{p \in \mathbf{K}_{s,d}} \left( end_{p,s} - start_{p,s} \right)}{| \mathbf{K}_{s,d} |}
\end{align}
where $\mathbf{K}_{s,d}$ is the set of all people who hop from job $s$ to job $d$, and $start_{p,s}$ and $end_{p,s}$  are the start and end dates of a job $s$ that a person $p$ takes, respectively. We note that the duration cost is computed based on the start/end dates of a source (instead of destination) job $s$, as our primary interest is a person's duration of stay at the source job before hopping to the destination job.

\item \textbf{Level gain} ($L_{s,d}$): This refers to the difference between the levels of two jobs $s$ and $d$, where the ``level'' of job $s$ is estimated by computing the \emph{average work experience} over all people who have job $s$ in their career trajectories.  \eepeng{Here, we assume that a job transition that gains much job level is favorable.  In our data, there is no explicit job level information.} The job level gain $L_{s,d}$ is thus defined by:
\begin{align}
\label{eqn:level_gain}
L_{s,d} &= L_{d} - L_{s} = \frac{\sum_{p \in |\mathbf{K}_d|} w_{p,d}}{|\mathbf{K}_d|}  - \frac{\sum_{p \in |\mathbf{K}_s|} w_{p,s} }{|\mathbf{K}_s|}
\end{align}
where $\mathbf{K}_s$ is the set of all people who ever took job $s$, and $w_{p,s}$ is the work experience of person $p$ at job $s$. he work experience is the duration since the last graduation date of a person till the time at which he/she finishes a particular job. Intuitively, a job with longer average work experience implies that a longer time is required to achieve that position, and so we can expect it to be a high-level job.
Examples of job with high level score $L_{s}$ according to our OPN data are ``Professor'' and ``Managing Director'', whereas examples of job with low $L_{s}$ are ``Intern'' and ``Teaching Assistant''.
Also note that a positive (negative) gain $L_{s,d}$ can be loosely viewed as a ``promotion'' (``demotion'')---in terms of the level of work experience required. We note that, although there is no ground truth available in our OPN data, our manual inspections show that the level gain provides a reasonable proxy for a promotion or demotion.

\item \textbf{Desirability gain} ($R_{s,d}$). This is defined as the difference between logarithm of the desirability of two jobs $s$ and $d$, where the ``desirability'' is computed using the \emph{PageRank} algorithm \cite{Page1999,Kapur:KDD2016}. In this case, we reinterpret the set of all job transitions $\mathcal{D}$ as a directed (unweighted) graph, where each node corresponds to a job $s$ and each edge a job transition $(s,d)$. PageRank views inbound edges as ``votes'', and the key idea is that ``votes'' from important nodes should carry more weight than ``votes'' from less important nodes \cite{Page1999}. As such, a job node that has high PageRank score reflects a ``desirable'' point where the flow of job transitions is heading to. We compute the desirability gain $R_{s,d}$ as:
\begin{align}
\label{eqn:desirability_gain}
R_{s,d} &= \log(P_{d}) - \log(P_{s})
\end{align}
where $P_{s}$ is the PageRank score of job $s$. In this work, we employ a \emph{weighted} version of PageRank \cite{Langville2005}, whereby the transition probabilities for each (source) node is proportional to the (out-)edge weights divided by the weighted out-degree of the node. In the context of this work, the weighted PageRank can be viewed as a measure of \emph{global competitiveness}. That is, a job with high PageRank reflects a ``desirable'' destination where the flow of job transitions is heading to. In this case, we use hop volume as the edge weight, based on the intuition that the volume matters in determining where the flow goes to. To prevent inflated PageRank due to sink nodes (i.e., nodes with zero out-degree), we also allow a jump to any random node in the graph with a probability of $\alpha$. Unless specified, we use $\alpha = 0.15$ by default in this work. 
\end{itemize}

Using the above criteria, we construct a three-dimensional payoff vector $\vec{f}(s,d) = [f_1(s,d), f_2(s,d), f_3(s,d)] = [-D_{s,d}, L_{s,d}, R_{s,d}]$. In contrast to $L_{s,d}$ and $R_{s,d}$, we use the negated value of $D_{s,d}$, because we view the duration as a cost to be minimized.

\section{Empirical Study}
\label{sec:experiments}

This section presents our empirical study on the dataset presented in Section~\ref{sec:dataset}.  We first describe the evaluation metrics and baseline methods used in our study. We then present our quantitative and qualitative analyses. Our analyses aim to demonstrate that one can optimize career paths starting from any job considering the different criteria altogether, and to illustrate some optimized career paths through some case examples.    


\subsection{Metrics and Baselines}
\label{sec:metrics}


\textbf{Combined metrics.} We evaluate for each actual path $P$ in our dataset in Figure~\ref{tab:basic_stats} (57,784 of them), the quality of optimized path returned by our proposed multicriteria utility method compared with the quality of optimized path returned by some baseline method combining all the different payoff criteria.  We define this path-specific metric based on \emph{product of improvement means} (PIM). The path-specific PIM value comparing the path returned by a method denoted by $opt$ for an actual path $P$ is defined as:
\begin{align}
\label{eqn:PIM_P}
\text{PIM}^\text{opt}_P = \min_{i} \text{sgn}(\mu_{P,i}) \times \prod_{i=1}^M |\mu_{P,i}|
\end{align}
where
\begin{align}
\mu_{P,i} = \sum_{(s,d) \in \mathbf{P}^\text{opt}(P, s_0)} f_i(s,d) - \sum_{(s,d) \in \mathbf{P}^\text{actual}(P, s_0)} f_i(s,d)
\end{align}
We then compare our method with a baseline by computing $\text{PIM}_P^\text{MUL/D}-\text{PIM}_P^\text{baseline}$.  This metric returns a positive value when our method returns a optimized path superior than that returned by the baseline method, and negative value otherwise.

\textbf{Baseline methods.} In this work, we evaluate our MUL/D-based \texttt{JobComposer} approach against several baselines. They consists of greedy based methods, methods that optimize single-criterion utility only for the purpose of finding the best performance one can achieve for each criterion, and a method assuming equal weights for different criteria. Specifically, the baseline methods include the following:
\begin{itemize}
\item \textbf{Greedy most common path}: Greedy path generation by following job transitions with the most number of people.
\item \textbf{Greedy shortest duration path}: Greedy path generation by following job transitions with the lowest $D_{s,d}$.
\item \textbf{Greedy level gain path}: Greedy path generation by following job transitions with the highest $L_{s,d}$.
\item \textbf{Greedy desirability gain path}: Greedy path generation by following job transitions with the highest $R_{s,d}$.
\item \textbf{Single-criterion utility path} ($-D_{s,d}$): Utility learning with negated $D_{s,d}$ as payoff criterion.
\item \textbf{Single-criterion utility path} ($L_{s,d}$): Utility learning with $L_{s,d}$ as payoff criterion.
\item \textbf{Single-criterion utility path} ($R_{s,d}$): Utility learning with $R_{s,d}$ as payoff criterion.
\item \textbf{Equally weighted utility path}: Utility learning via linear scalarization of payoff criteria: $w_1 L_{s,d} - w_2 D_{s,d} + w_3 R_{s,d}$
\end{itemize}

\eepeng{The greedy methods are designed to select the job that maximizes some single criterion at every new job transition.  Since they construct the overall path by making incremental decisions, they may not return the optimal path even for the criterion they try to optimize.  The single criterion methods can address this shortcoming, but do not optimize for multiple criteria.  The multi-criteria utility method to be compared with ours is the Equally weighted utility path method.}

\begin{figure}[!t]
\centering
\includegraphics[width=1.0\columnwidth]{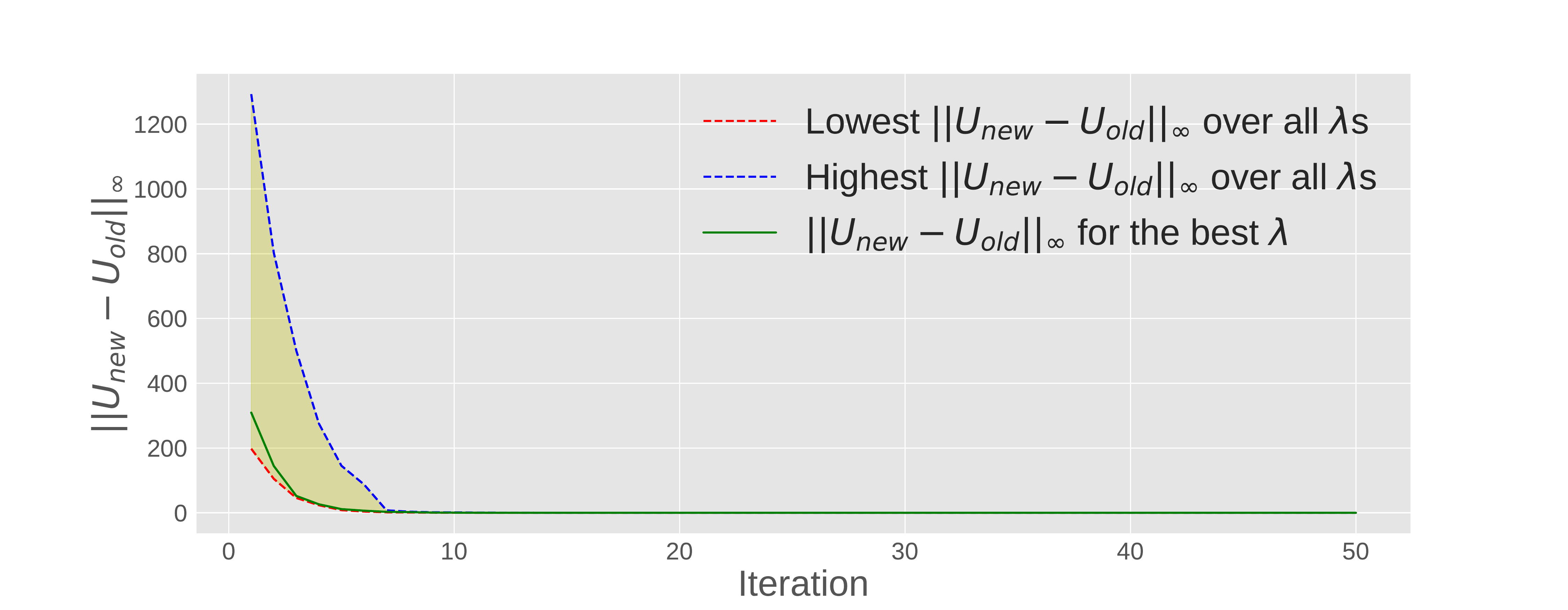}
\caption{Convergence of our multicriteria utility method}
\label{fig:utility_convergence}
\end{figure}

\textbf{Parameter settings.}  \richard{For all the utility-based baselines (including ours), we set the discount factor and maximum iteration as $\gamma = 0.7$ and $T_{max} = 50$, respectively, which give the best results overall. For the equally weighted utility method, as we do not have one preferred criterion , we equalize the weights as: $w_1 = w_2 = 1$ and $w_3 = \beta = 500$. We set an equal weight for $D_{s,d}$ and $L_{s,d}$, since they have the same unit (i.e., months). Meanwhile, we use $w_3 = \beta$ to bring $R_{s,d}$ after  to (roughly) the same scale. To test for the statistical significance of our method relative to the baselines, we use the \emph{Wilcoxon signed-rank test} \cite{Wilcoxon1945}, which is a non-parametric statistical significance test for two paired samples, which cannot be assumed to be normally distributed. A low $p$-value below a threshold (typically 0.05 or 0.01) indicates that they are significantly different.}

\subsection{Quantitative Analysis}
\label{sec:quantitative}

\textbf{Convergence analysis.} We firstly examine the convergence property of our MUL/D algorithm. In particular, We would like to show that our MUL/D method converges within a small number of iterations.  The shaded region in Figure \ref{fig:utility_convergence} shows the evolution of the $L_{\infty}$-norm of the utility changes between two successive iterations $|| U^{(k+1)} - U^{(k)}||_{\infty}$ of all possible weights in $\bm{\lambda}$ over iterations. The solid curve in the shaded region represents the $L_{\infty}$-norm of $\lambda^*$ as mentioned in (\ref{eqn:PIM}), whereas the lower and upper dashed curves represent minimum and maximum $L_{\infty}$-norm respectively. In terms of utility changes, we can see that our method converges to a stationary point within $10$ iterations, regardless of the choice of $\lambda$. This provides an empirical justification for the convergence trait as per (\ref{eqn:convergence}).

\begin{figure*}[!t]
\centering
\includegraphics[width=1.0\textwidth]{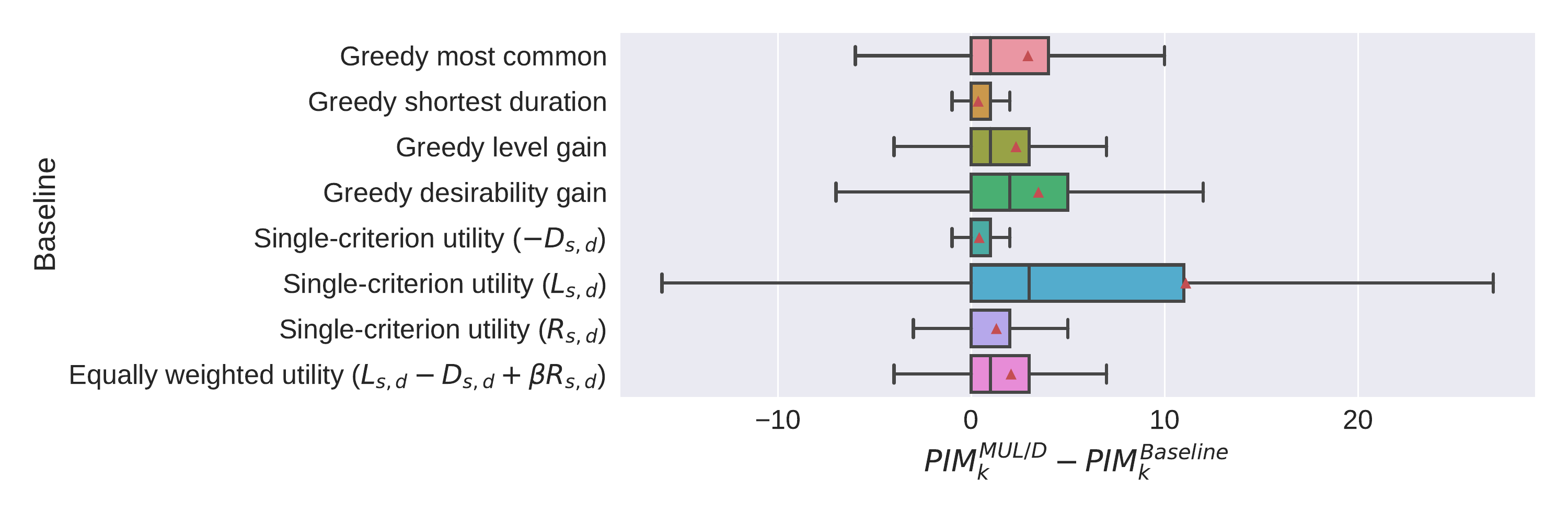}
\caption{Path-based combined metric improvements for different baseline methods}
\label{fig:pathPIM_boxplot}
\end{figure*}


\textbf{Comparison with baselines.} 
First, we show the mean and median of $PIM_P^\text{MUL/D}-PIM_P^\text{Baseline}$ of actual paths when they are optimized by our method and some baseline method in Figure~\ref{fig:pathPIM_boxplot}. 
The results show that our method outperforms all the baselines for most of the actual paths as $PIM_P^\text{MUL/D}-PIM_P^\text{Baseline}$ values are largely positive. In particular, our method is very much better than single-criterion utility method based on job level $L$, greedy desirability gain method and greedy most common method. 

\begin{table*}[!t]
\scriptsize
\centering
\caption{Benchmarking results of different path planning methods}
\label{tab:benchmark}
\begin{tabular}{|l|c|c|c|c|c|c|c|}
\hline
\textbf{Method} & \textbf{PIM} & \multicolumn{2}{|c|}{$D^{\text{actual}} - D^{\text{optimized}}$} & \multicolumn{2}{|c|}{$L^{\text{optimized}} - L^{\text{actual}}$} & \multicolumn{2}{|c|}{$R^{\text{optimized}} - R^{\text{actual}}$}  \\
\cline{3-8}
  									&  	& \textbf{Mean} &  \textbf{$p$-value} &   \textbf{Mean} &  \textbf{$p$-value} &  \textbf{Mean} &    \textbf{$p$-value} \\
\hline
Greedy most common       			&  -74.01 & -88.80 &    0.00e+00 (-) &  11.67 &  0.00e+00 (-) &  -0.07 & 0.00e+00 (-)  	\\
Greedy shortest duration 			&  -19.46 	&  23.06 &    0.00e+00 (+) &   2.67 &  0.00e+00 (-) &  -0.32 & 0.00e+00 (-) \\
Greedy level gain        			&  -255.37 	& -28.39 &    0.00e+00 (-) &  26.84 &  0.00e+00 (+) &  -0.34 & 0.00e+00 (-) \\
Greedy desirability gain 			&  -1057.25	& -57.38 &    0.00e+00 (-) &  19.19 &  3.34e-04 (-) &   0.96 & 0.00e+00 (+) \\
Single-criterion utility ($-D_{s,d}$) &  -198.45 &  61.30 &  0.00e+00 (+) &  -6.55 &  0.00e+00 (-) &  -0.49 & 0.00e+00 (-) 	\\
Single-criterion utility ($L_{s,d}$) &  -2478.66 & -90.45 &   0.00e+00 (-) &  81.52 &  0.00e+00 (+) &  -0.34 & 0.00e+00 (-) \\
Single-criterion utility ($R_{s,d}$) &  -240.31 	& -36.65 &   0.00e+00 (-) &   5.96 &  0.00e+00 (-) &   1.10 & 0.00e+00 (+) \\
Equally weighted utility 			&  -577.01 	& -42.83 &    0.00e+00 (-) &  12.31 &  0.00e+00 (-) &   1.09 & 0.00e+00 (+) \\
Multicriteria utility 				&   241.73 	&  16.36 &         - 	   &  19.51 &      - 		&   0.76 &     - 		\\
\hline
\multicolumn{8}{l}{(+): significantly better than multicriteria utility; (-): significantly worse than multicriteria utility}
\end{tabular}
\end{table*}

\begin{table*}[!t]
\scriptsize
\centering
\caption{Examples of career path optimization results}
\label{tab:examples}
\begin{tabular}{|l|l|}
\hline
\#1 & \textbf{Actual path}: (Telecom., 10,001+ employees, Technical Officer) $\rightarrow$ (Telecom., 5001-10K employees, Engineer) \\
& $\rightarrow$ (Telecom., 10,001+ employees, Associate Eng.) \\
\cline{2-2}
   & \textbf{Greedy most common path}:  (Telecom., 10,001+ employees, Tech. Officer) \\
  & $\rightarrow$ (International Trade \& Development, 5001-10,000 employees, Tech. Manager)  \\
\cline{2-2}
   & \textbf{Multicriteria utility path}: (Telecom., 10,001+ employees, Tech. Officer) $\rightarrow$ (Telecom., 5001-10K employees, Eng.) \\
   & $\rightarrow$ (Telecom., 5001-10K employees, Project Leader) \\
\cline{2-2}
	& \textbf{Metric improvement for greedy most common path}: $D^{\text{actual}} - D^{\text{greedy most common}} =$ -72.00,\\ & $L^{\text{greedy most common}} - L^{\text{actual}} =$ 107.15, $R^{\text{greedy most common}} - R^{\text{actual}} =$ -0.06 \\
\cline{2-2}
    & \textbf{Metric improvement for multicriteria utility path}: $D^{\text{actual}} - D^{\text{multicriteria utility}} =$ 70.00, \\
    & $L^{\text{multicriteria utility}} - L^{\text{actual}} =$ 100.73, $R^{\text{multicriteria utility}} - R^{\text{actual}} =$ 0.04 \\
\hline
\#2 & \textbf{Actual path}: (Banking, 10,000+ employees, Analyst) $\rightarrow$ (IT \& Services, 5001-10,000 employees, Analyst) \\
& $\rightarrow$ (IT \& Services, 5,000 - 10,000 employees, Team Leader) $\rightarrow$ (Banking, 10,000+  employees, Credit Risk Analyst) \\
\cline{2-2}
   & \textbf{Greedy most common path}:  (Banking, 10,000+ employees, Analyst) $\rightarrow$ (Banking, 10,000+ employees, Associate) \\
  & $\rightarrow$ (Banking, 10,000+ employees, Assistant Vice President)  $\rightarrow$ (Banking, 10,000+ employees, Vice President) \\
 & $\rightarrow$ (Banking, 10,000+ employees, Director) $\rightarrow$ (Financial Services, 10,000+ employees, Executive Director) \\
   & $\rightarrow$
   (Banking, 10,000+ employees, Managing Director)  \\
\cline{2-2}
   & \textbf{Multicriteria utility path}: (Banking, 10,000+ employees, Analyst) $\rightarrow$ (Banking, 10,000+ employees, Consultant) \\
   & $\rightarrow$ (IT \& Services, 10,000+ employees, Technical Consultant) \\
   & $\rightarrow$ (IT \& Services, 10,000+ employees, Project Manager) \\
   & $\rightarrow$ (Financial Services, 10,000+ employees, Assistant Vice President) \\
\cline{2-2}
	& \textbf{Metric improvement for greedy most common path}: $D^{\text{actual}} - D^{\text{greedy most common}} =$ -165.24,\\  & $L^{\text{greedy most common}} - L^{\text{actual}} =$ 160.56, $R^{\text{greedy most common}} - R^{\text{actual}} =$ 0.66 \\
\cline{2-2}
    & \textbf{Metric improvement for multicriteria utility path}: $D^{\text{actual}} - D^{\text{multicriteria utility}} =$ 56.00, \\
    & $L^{\text{multicriteria utility}} - L^{\text{actual}} =$ 78.89,  $R^{\text{multicriteria utility}} - R^{\text{actual}} =$ 1.09 \\
\hline
\#3 & \textbf{Actual path}: (Semiconductors, 10,001+ employees, Process Eng.) $\rightarrow$ (Semiconductors, 201-500 employees, Process Eng.) \\
& $\rightarrow$ (Semiconductors, 5001-10,000 employees, Equip. Eng.) $\rightarrow$ (Semiconductors, 201-500 employees, Field Appln. Eng.) \\
\cline{2-2}
   & \textbf{Greedy most common path}:  (Semiconductors, 10,001+ employees, Process Engineer) \\
   & $\rightarrow$ (Semiconductors, 10,001+ employees, Senior Process Eng.)  \\
\cline{2-2}
   & \textbf{Multicriteria utility path}: (Semiconductors, 10,001+ employees, Process Eng.) \\
   & $\rightarrow$ (Banking, 10,001+ employees, Rel. Manager) $\rightarrow$ (Banking, 10,001+ employees, Associate Director) \\ & $\rightarrow$
   (Banking, 10,001+ employees, Managing Director) \\
\cline{2-2}
	& \textbf{Metric improvement for greedy most common path}: $D^{\text{actual}} - D^{\text{greedy most common}} =$ 79.75,\\ 
    & $ L^{\text{greedy most common}} - L^{\text{actual}} =$ 21.68, $R^{\text{greedy most common}} - R^{\text{actual}} =$ 0.59 \\
\cline{2-2}
    & \textbf{Metric improvement for multicriteria utility path}: $D^{\text{actual}} - D^{\text{multicriteria utility}} =$ 67.42, \\
    & $L^{\text{multicriteria utility}} - L^{\text{actual}} =$ 105.44, $R^{\text{multicriteria utility}} - R^{\text{actual}} =$ 1.39 \\
\hline
\multicolumn{2}{l}{\textbf{Note}: The format of a job is $(i,c,j)$, where $i$ is the industry code, $c$ is the company size category, and $j$ is the job title.
}
\end{tabular}
\end{table*}

Next, Table~\ref{tab:benchmark} summarizes how well our method fares against the baselines according to the improvement in combined criteria measured by $PIM_P^\text{optimized}-PIM_P^\text{actual}$, and in single criteria (i.e., duration cost, level gain, or desirability gain) of the optimized career path returned by the method against that of the observed career path.   $PIM^{\text{optimized}}$ and $PIM^{\text{actual}}$  denote the PIM value of optimized path and actual path respectively. For single criteria such as duration cost. The duration cost of a path is defined as $D_{\mathbf{P}} = \sum_{(s,d) \in \mathbf{P}} D_{s,d}$, and $D_P^\text{actual}-D_P^\text{optimized}$ denotes the improvement in duration cost.  Other single criteria-based improvement can be defined in a similar manner.  
As shown in Table~\ref{tab:benchmark}, our MUL/D approach returns the best combined performance over different criteria measured by the average improvement of PIM.  It is the only method that return positive PIM value.  It is also the only method that returns positive mean and median improvement in every criteria.  
The greedy and single-criterion methods tend to do well only on one specific criterion, but not the others.  Hence, they could return positive gains in one criterion but negative ones in other criteria.  

Compared to the equally weighted utility method, our MUL/D approach gives better mean for $PIM_P^\text{optimized}-PIM_P^\text{actual}$, $D^{\text{actual}} - D^{\text{optimized}}$ and $L^{\text{optimized}} - L^{\text{actual}}$, though the former is superior in terms of $R^{\text{optimized}} - R^{\text{actual}}$.   This result is reasonable as the equally weighted utility method is a reasonably strong baseline.
Table \ref{tab:benchmark} also includes the two-sided $p$-values of the Wilcoxon test. It is evident that all $p$-values are very small (i.e., $\ll 0.01$). Hence, the performance differences between our method with the other baselines are statistically significant. While above mentioned baselines are conceptually weaker, they are not necessarily easy to beat in practice. 

\textbf{Runtime analysis.} In the above quantitative analysis, MUL/D required less than one and a half hour to complete the analysis of 265,000+ job transitions.  With parallel computation, the time required can be further reduced significantly. We nevertheless leave this detailed runtime analysis to future work.

\subsection{Qualitative Analysis}
\label{sec:qualitative}

\richard{We also study a few career path examples in detail in order to qualitatively evaluate the career path recommendations made by \texttt{JobComposer}. Specifically, we look at examples whereby $D^{\text{actual}} - D^{\text{multicriteria utility}} \geq 50$ months, $L^{\text{multicriteria}} - L^{\text{actual}} \geq 70$ months, and $R^{\text{multicriteria utility}} - R^{\text{actual}} \geq 0$.  These are the examples that can show significant improvement according to our chosen multiple criteria.  Table \ref{tab:examples} shows three such examples. For comparison, we also show the corresponding recommendations made by the greedy most common path method. It can be seen that our multicriteria utility learning approach yields better metric improvements (at least two metrics) in comparison to the greedy most common path method.}

\richard{It is interesting see how our method can devise optimized paths that are sensible and yet yield high improvements. The first examples illustrates a person who begins his career as a Technical Officer in Telecommunications Industry with a large company having more than 10,000 employees and remains in the same industry for entry-level engineering roles with similar company sizes. In this case, though the greedy most common method provides a higher level gain by switching to a lower company size in the Internet Industry, it takes 72 additional months and the desirability gain is less. In contrast, our method suggests to remain in the same Telecommunications Industry but in a relatively smaller company (5000-10,000 employees) ending up as a project leader, saving 70 months and having higher desirability gain.}

\richard{In the second example, an Analyst in a very large bank, following the path suggested by our method (which advises to not immediately switch to a different Industry) saves 56 months and will end up as an Assistant Vice President in Financial Services Industry (higher level and desirability gains). Though the path suggested by greedy most common method has much higher level gain, the person has to wait for around 165 months to achieve it and also the desirability gain is much lesser than that of our method.}

\richard{The last example shows how a person working in a large Semiconductor company can improve his career by switching to Banking Industry as a Relationship Manager, ending up as a Managing Director of a large bank (higher level and desirability gain) and saves around 67 months. Following the greedy most common method, the person instead of switching to different smaller companies in the same industry, could get a promotion in the same industry of same company size, saving 80 months. Comparing to our method, the greedy most common method has slightly better duration gain, but it has much lesser level and desirability gains.}

\richard{In sum, our qualitative study shows that \texttt{JobComposer} produces sensible career path recommendations, with better quality than those of the greedy most common path, also justifying our earlier hypothesis that the common paths may not be always optimal and can be improved. Nevertheless, it must be noted that all the paths recommended in Table \ref{tab:examples} only show what is possible, but do not necessarily guarantee that the next job in the path can always be achieved.  Constraining the path to available openings and estimating the probability of successfully obtaining a career is beyond the scope of this paper, but is certainly the direction of our future work.}



\section{Conclusion}
\label{sec:conclusion}

In this paper, we put forward a new data-driven approach for automated career path planning and optimization called \texttt{JobComposer}. To realize \texttt{JobComposer}, we first formulate career path optimization as a utility learning problem on top of a Markov process. Based on this formulation, we then develop a novel decomposition-based multicriteria utility learning and multicriteria utility selection procedures to devise a career path that optimizes multiple criteria simultaneously and identify the best tradeoff among different criteria, respectively. Comprehensive quantitative and qualitative studies using a city state-based OPN data have demonstrated the efficacy of our approach.

While \texttt{JobComposer} offers a powerful methodology for career path planning, there remains room of improvements. To endow a greater level of personalization within its career optimization processes, we would like to develop a more advanced method that utilizes a richer set of user features (e.g., skills, education level, etc.). Additionally, we plan to augment relevant auxiliary data---such as employees' salaries---and conduct studies with a more comprehensive set of payoff criteria.

%
%
%
%
\bibliography{references}
\bibliographystyle{splncs04}

\end{document}